\documentclass[aps,prb,twocolumn,amsmath,amssymb,showpacs,superscriptaddress]{revtex4}

\usepackage{graphicx}
\usepackage{dcolumn}
\usepackage{bm}

\begin{document}

\title{Surface polaritons on metallic and semiconducting cylinders: \\
A complex angular momentum analysis}

\author{St\'ephane Ancey}
\email{ancey@univ-corse.fr}
\affiliation{ UMR CNRS 6134 SPE, Equipe Ondes et Acoustique, \\
Universit\'e de Corse, Facult\'e des Sciences, Bo{\^\i}te Postale
52, 20250 Corte, France}

\author{Yves D\'ecanini}
\email{decanini@univ-corse.fr}
\affiliation{ UMR CNRS 6134 SPE,
Equipe Physique Semi-Classique (et) de la Mati\`ere Condens\'ee,
\\ Universit\'e de Corse, Facult\'e des Sciences, Bo{\^\i}te
Postale 52, 20250 Corte, France}

\author{Antoine Folacci}
\email{folacci@univ-corse.fr}
\affiliation{ UMR CNRS 6134 SPE,
Equipe Physique Semi-Classique (et) de la Mati\`ere Condens\'ee,
\\ Universit\'e de Corse, Facult\'e des Sciences, Bo{\^\i}te
Postale 52, 20250 Corte, France}

\author{Paul Gabrielli}
\email{gabrieli@univ-corse.fr}
\affiliation{ UMR CNRS 6134 SPE,
Equipe Ondes et Acoustique,
\\ Universit\'e de Corse, Facult\'e des Sciences, Bo{\^\i}te
Postale 52, 20250 Corte, France}

\date{\today}

\begin{abstract}

We revisit scattering of electromagnetic waves from metallic and
semiconducting cylinders in the framework of complex angular
momentum techniques. We prove that ``resonant surface polariton
modes" are generated by a unique surface wave, i.e. a surface
polariton, propagating close to the cylinder surface. This surface
polariton corresponds to a particular Regge pole of the $S$-matrix
of the cylinder. From the associated Regge trajectory we can
construct semiclassically the spectrum of the complex frequencies
of the resonant surface polariton modes which can be considered as
Breit-Wigner-type resonances. Furthermore, by taking into account
Stokes' phenomenon, we derive an asymptotic expression for the
position in the complex angular momentum plane of the surface
polariton Regge pole. We then describe semiclassically the surface
polariton and provide analytical expressions for its the
dispersion relation and its damping. All these features allow us
to consider the photon-cylinder system as a kind of artificial
atom where the photon plays the role of the electron. Finally, we
briefly discuss the implication of our results for two-dimensional
photonic crystals.

\end{abstract}

\pacs{78.20.-e, 41.20.Jb, 73.20.Mf, 42.25.Fx}

\maketitle

\section{Introduction}

In recent years, since the work of McGurn and Maradudin
\cite{McGurnMaradudin93}, there has been a growing interest in the
study of photonic crystals containing metallic or semiconducting
components. The frequency-dependence of the dielectric function
$\epsilon (\omega)$ of these materials, as well as the presence of
a frequency range where $\epsilon (\omega) <0$, lead to new
features which do not exist in conventional photonic crystals and
which are mainly linked to the presence of surface polaritons
(SP's). With this in mind, we intend to revisit the theory of SP's
propagating close to curved metal-dielectric or
semiconductor-dielectric interfaces. In this paper, we will
emphasize the case of the circular cylindrical interface. For
reviews on photonic crystal physics see
Refs.~\onlinecite{Joan1er,Joan2eme,Sakoda2001}. Among the numerous
articles dealing with SP's in photonic crystals, we refer to
Refs.~\onlinecite{McGurnMaradudin93,KusmiakMP1994,SigalasCHS1995,
ZhangHLXM1996, KusmiakMG1997,KusmiakM1997,Moroz2000, Sakoda2001b,
MorenoEH2002,OchiaiSD2002}, i.e. to a non-exhaustive list of works
dealing more particularly with two-dimensional arrays of cylinders
of circular cross section for which SP's may be excited and which
are consequently relevant to our study.

SP's supported by flat metal-dielectric or
semiconductor-dielectric interfaces can be easily described from a
theoretical point of view and their properties can be obtained
from rather elementary calculations involving homogeneous and
inhomogeneous plane waves. It should be noted that in this
context, an SP is clearly defined as a surface wave propagating
close to the interface with an amplitude that decays in an
exponential fashion perpendicularly to the interface and into both
medium. By contrast, in the presence of cylindrical (this is also
true for spherical) interfaces, the corresponding theoretical
analysis is a little bit more complicated (see, for example,
Refs.~\onlinecite{EngRup68,PfeifferEcoNgai74,AshleyEm74,MartinosEc83}):
Bessel functions must be introduced and transcendental equations
involving them must be numerically solved, for example, in order
to understand resonance phenomenons associated with SP's. As a
consequence, a clear physical description of scattering of
electromagnetic waves from curved metal-dielectric or
semiconductor-dielectric interfaces does not still exist. In the
scientific literature, this fact sometimes leads to a semantic
ambiguity with the expression ``surface polaritons" denoting the
surface waves propagating close to the interface as well as the
resonant electromagnetic modes they generate. In order to avoid
such an ambiguity and to distinguish between the two physical
phenomenons, we shall use the more appropriate expression
``resonant surface polariton modes" (RSPM's) for the latter.

Since the sixties, mainly under the impetus of Nussenzveig,
asymptotic (i.e., semiclassical) techniques which use analytic
continuation of partial-wave representations (Mie sums) have been
developed to understand scattering of electromagnetic waves from
dielectric objects. Together these techniques form the complex
angular momentum (CAM) method. In electromagnetism, it can be
considered as a refinement of ray optics which takes into account
``tunneling aspects" of scattering and therefore includes
diffractive rays associated with surface waves. Of course, the CAM
method is an asymptotic approach and formally it is only valid at
length scales ``large" compared to the wavelength of the
electromagnetic field.

The CAM  method originates from the pioneering work of Watson
\cite{Watson18} dealing with the propagation and diffraction of
radio waves around the earth. It has since been successfully
introduced in various domains of physics. The success of the CAM
method is mainly due to its ability to provide a clear description
of a given scattering problem by extracting the physical
information (linked to the geometrical and diffractive aspects of
the scattering process) which is hidden in partial-wave
representations and then to semiclassically describe resonance
phenomenons. Here the dual structure of the $S$-matrix associated
with a given scattering problem plays a crucial role. Indeed, the
$S$-matrix is a function of both the frequency $\omega$ and the
angular momentum index $\ell$. It can be analytically extended
into the complex $\omega$-plane as well as into the complex
$\ell$-plane (CAM plane). Its poles lying in the fourth quadrant
of the complex $\omega$-plane are the complex frequencies of the
resonant modes. In other words, the behavior of the $S$-matrix in
the complex $\omega$-plane permits us to investigate resonance
phenomenons. The structure of the $S$-matrix in the complex
$\ell$-plane allows us, by using integration contour deformations,
Cauchy's Theorem and asymptotic analysis, to provide a
semiclassical description of scattering in terms of rays. In that
context, the poles of the $S$-matrix lying in the CAM plane (the
so-called Regge poles) are associated with diffraction. Of course,
when a connection between these two faces of the $S$-matrix can be
established, resonance aspects of scattering are then
semiclassically interpreted. For reviews of the CAM method we
refer to the monographs of Newton \cite{New82}, Nussenzveig
\cite{Nus92} and Grandy \cite{Grandy} as well as to references
therein for various applications in quantum mechanics, nuclear
physics, electromagnetism, optics, acoustics and seismology. For
recent applications in more ``exotic" contexts, we refer to
Refs.~\onlinecite{Andersson1,Andersson2,DecaFJPRD} where a CAM
analysis of black hole scattering and black hole gravitational
radiation is provided and to Ref.~\onlinecite{DecaFPRA} where the
Aharonov-Bohm effect is considered.

As far as we know, the CAM method has never been used to
understand scattering of electromagnetic waves from metallic and
semiconducting objects. In fact, because of the
frequency-dependence of the dielectric function of metals and
semiconductors, the extension of the ideas of Nussenzveig and
coworkers\cite{Nus92} to such a problem is not quite as obvious as
it seems at first sight. In this article, we make some steps in
that direction but with a rather modest goal. Indeed, we only
consider the scattering of TE waves ($H$ polarization) by a
metallic or semiconducting circular cylinder surrounded by a
dielectric medium. We limit our study to that case because SP's
are not excited in the $E$-polarization configuration. From the
$S$-matrix of the cylinder, using CAM techniques, we develop a
semiclassical description of the scattering aspects linked to
SP's. More precisely, we prove that RSPM's are generated by a
unique SP propagating close to the cylinder surface. This surface
wave is associated with a particular Regge pole of the $S$-matrix
of the cylinder. From the corresponding Regge trajectory, i.e.
from the curve traced out in the CAM plane by this Regge pole as a
function of the frequency, we can construct semiclassically the
spectrum of the complex frequencies of RSPM's which can be
considered as Breit-Wigner-type resonances. Furthermore, by
carefully taking into account Stokes phenomenon, we derive an
asymptotic expression for the position of the SP Regge pole in the
CAM plane and then, we can describe semiclassically the SP. In
some sense, our results allow us to consider the photon-cylinder
system as an artificial atom: RSPM's are long-lived quasibound
states for this atom while the trajectory of the SP which
generates them and which is supported by the cylinder surface is a
Bohr-Sommerfeld-type orbit.

Our paper is organized as follows. In Section 2, we introduce our
notations and we construct the $S$-matrix of the system. We then
discuss the resonant aspects of our problem. In Section 3, by
using CAM techniques, we establish the connection between the SP
propagating close to the surface cylinder and the associated
RSPM's. In Section 4, we describe semiclassically the SP by
providing analytic expressions for its dispersion relation and its
damping. We then deduce analytic approximations for the excitation
frequencies of RSPM's. Finally, in Section 5, we conclude our
article by considering some possible extensions of our work and by
briefly discussing the implication of our results in the context
of two-dimensional photonic crystal physics.

\section{Exact $S$-matrix and scattering resonances}

From now on, we consider the scattering of an electromagnetic wave
by a metallic or semiconducting circular cylinder with a
frequency-dependent dielectric function $\epsilon_c (\omega)$
which is embedded in a host medium of infinite extent with
constant dielectric function $\epsilon_h$ (region I). In the usual
cylindrical polar coordinate system $(\rho ,\theta ,z)$ the
cylinder occupies a region corresponding to the range $0 \le \rho
< a$ (region II). We also assume that the magnetic field
$\mathbf{H}$ is parallel to the axis of the cylinder ($H$
polarization) and we choose to treat our problem in a
two-dimensional setting, ignoring the $z$ coordinate. Furthermore,
in the following, we implicitly assume the time dependence
$\exp(-i\omega t)$ for the magnetic field and we shall sometimes
use the wave numbers
\begin{equation}
k^{\mathrm {I}}(\omega)=\left(\frac{\omega }{c} \right)
\sqrt{\epsilon_h} \quad \mathrm{and} \quad k^{\mathrm
{II}}(\omega)=\left(\frac{\omega }{c} \right) \sqrt{ \epsilon_c
(\omega)}.
\end{equation}
Here $c$ is the velocity of light in vacuum.

As far as the dielectric function of the cylinder is concerned, we
assume it presents a Drude-like behavior
\cite{AshcroftMermin,MarkFox}
\begin{equation}
\epsilon_c (\omega)= \epsilon_\infty \left( 1-
\frac{\omega_p^2}{\omega ^2} \right), \label{PermDrude}
\end{equation}
or the ionic crystal behavior \cite{AshcroftMermin,MarkFox}
\begin{equation}
\epsilon_c (\omega)=  \epsilon_\infty \left( \frac{\omega_L^2 -
\omega^2 }{\omega_T^2 -  \omega^2 } \right). \label{PermCristIon}
\end{equation}
In both cases, $\epsilon_\infty$ is the high-frequency limit of
the dielectric function. In Eq.~(\ref{PermDrude}), $\omega_p$ is
the plasma frequency. In Eq.~(\ref{PermCristIon}), $\omega_T$ and
$\omega_L$ respectively denote the transverse-optical-phonon
frequency and the longitudinal-optical-phonon frequency. In the
first case, SP's follow from the coupling of the electromagnetic
wave with the plasma wave and are usually called surface plasmon
polaritons. In the second one, SP's follow from the coupling of
the electromagnetic wave with the longitudinal and transverse
acoustic waves and are usually called surface phonon polaritons.
Eq.~(\ref{PermDrude}) can be used to describe the dielectric
behavior of certain metals and semiconductors (Si, Ge, InSb,
$\dots$ ) while Eq.~(\ref{PermCristIon}) can be used to
investigate the optical properties of other semiconductors such as
GaAs.

From Maxwell's equations it is easy to show that the $z$-component
of the magnetic field satisfies the Helmholtz equation
\begin{subequations}\label{HEqu}
\begin{eqnarray}&& \left[ \Delta_{\bf x} +\left(
\frac{\omega}{c} \right)^2 \epsilon_c (\omega) \right]
H_z^{\mathrm {II}} ({\bf x}) =0  \quad \mathrm{for} \  0  \le \rho
< a,
\nonumber \\
&&    \label{HEqu1} \\
&& \left[ \Delta_{\bf x} +\left( \frac{\omega}{c}
\right)^2\epsilon_h \right] H_z^{\mathrm {I}} ({\bf x}) =0 \quad
\mathrm{for} \ \rho
> a,  \label{HEqu2}
\end{eqnarray}
\end{subequations}
where ${\bf x}=(\rho,\theta)$ and with the Laplacian $\Delta_{\bf
x}$ given, in the polar coordinate system, by
\begin{equation}\label{lapl}
\Delta_{\bf x} = \frac{\partial ^2}{\partial
\rho^2}+\frac{1}{\rho}\frac{\partial }{\partial
\rho}+\frac{1}{\rho^2} \frac{\partial ^2}{\partial \theta^2}.
\end{equation}
Furthermore, from the continuity of the tangential components of
the electric and magnetic fields (i.e., of $E_\theta$ and $H_z$)
at the interface between regions I and II, it can be shown that
the $z$-component of the magnetic field satisfies for
$0\leq\theta<2\pi$
\begin{subequations}\label{BCHz}
\begin{eqnarray}
&&H_z^{\mathrm I}(\rho =a,\theta)=H_z^{\mathrm {II}}(\rho
=a,\theta) , \label{BCHz1} \\
&& \frac{1}{\epsilon_h} \, \frac{\partial H_z^{\mathrm
I}}{\partial \rho}(\rho =a,\theta)=\frac{1}{\epsilon_c (\omega)}
\, \frac{\partial H_z^{\mathrm {II}}}{\partial \rho}(\rho
=a,\theta).  \label{BCHz2}
\end{eqnarray}
\end{subequations}

We are first interested in the construction of the $S$-matrix for
the cylinder. Because of the cylindrical symmetry of the
scatterer, the $S$-matrix is diagonal and its elements $S_{\ell
\ell'}$ are given by $S_{\ell \ell'}=S_\ell \ \delta _{\ell
\ell'}$. For a given angular momentum index $\ell \in \mathbf{Z}$,
the coefficient $S_\ell$ is obtained from the partial wave
${(H_z)}_\ell$ solution of the following problem (here we extend,
{\it mutatis mutandis}, the quantum mechanical approach developed
in Ref.~\onlinecite{Mott65}):
\begin{description}
  \item (i) ${(H_z)}_\ell$ satisfies the Helmholtz equation (\ref{HEqu}),
  \item (ii) ${(H_z)}_\ell$ satisfies the boundary conditions (\ref{BCHz}),
  \item (iii) at large distance, ${(H_z)}_\ell$ has the asymptotic behavior
\begin{eqnarray}
&&{(H_z)}_\ell (\rho,\theta) \underset{\rho \to +\infty}{\sim}
\frac{1}{\sqrt{2\pi k^{\mathrm {I}}\rho}}
\left(e^{-i(k^{\mathrm {I}}\rho -\ell\pi/2-\pi/4)} \right. \nonumber \\
&& \qquad\qquad\qquad +   \left. S_\ell (\omega)  e^{i(k^{\mathrm
{I}}\rho -\ell\pi/2-\pi/4)}\right)e^{i\ell\theta}. \nonumber
\end{eqnarray}
\end{description}
Outside the cylinder (region I), the solution of (\ref{HEqu}) is
expressible in terms of Bessel functions (see
Ref.~\onlinecite{AS65}) as a linear combination of
$J_{\ell}(k^{\mathrm {I}}\rho)e^{i\ell\theta}$ and $H^{(1)}_{\ell
}(k^{\mathrm {I}}\rho)e^{i\ell\theta}$. Inside the cylinder
(region II), it is proportional to $J_\ell(k^{\mathrm
{II}}\rho)e^{i\ell\theta}$. As a consequence, the partial wave
${(H_z)}_\ell$  solution of (i) and (ii) can be obtained exactly.
Then, by using the standard asymptotic behavior of Hankel
functions for $x \to \infty$ (see Ref.~\onlinecite{AS65})
\begin{subequations}\label{AsympHank}
\begin{eqnarray}
&&H_\ell^{(1)}(x)  {\sim} \sqrt{2/(\pi x)} e^{i(x-\ell \pi /2 -\pi
/4)},  \label{AsympHank1} \\
&& H_\ell^{(2)}(x)  {\sim} \sqrt{2/(\pi x)} e^{-i(x-\ell \pi /2
-\pi /4)},  \label{AsympHank2}
\end{eqnarray}
\end{subequations}
we find from (iii) the expression of the diagonal elements
$S_\ell$ of the $S$-matrix. We have
\begin{equation}\label{ExprMS}
S_\ell(\omega)=1-2\frac{D^{(1)}_\ell(\omega)}{D_\ell(\omega)}
\end{equation}
where $D^{(1)}_\ell(\omega)$ and $D_\ell(\omega)$ are two $2\times
2$ determinants which are explicitly given by
\begin{subequations}\label{DD}
\begin{eqnarray}
&D^{(1)}_\ell(\omega)=k^{\mathrm {II}}(\omega) J'_\ell (k^{\mathrm
{I}}(\omega)a) J_\ell(k^{\mathrm {II}}(\omega)a) \nonumber \\
& \quad -k^{\mathrm {I}}(\omega) J_\ell(k^{\mathrm
{I}}(\omega)a)J'_\ell(k^{\mathrm
{II}}(\omega)a),  \label{DDa}\\
&D_\ell(\omega)=k^{\mathrm {II}}(\omega) H_\ell^{(1)'} (k^{\mathrm
{I}}(\omega)a) J_\ell(k^{\mathrm {II}}(\omega)a) \nonumber \\
& \quad -k^{\mathrm {I}}(\omega) H_\ell^{(1)} (k^{\mathrm
{I}}(\omega)a)J'_\ell(k^{\mathrm {II}}(\omega)a). \label{DDb}
\end{eqnarray}
\end{subequations}
The unitarity of the $S$-matrix \cite{New82}, which expresses the
energy conservation, can be easily verified from (\ref{ExprMS})
and (\ref{DD}) by using elementary properties of Bessel functions.
The reciprocity property \cite{New82}, which is associated with
time-reversal invariance, is also satisfied because $S_\ell$ is an
even function of $\ell$.

The $S$-matrix is of fundamental importance because it contains
all the information about the scattering process. Its components
appear in the Green functions of the problem, in the scattered
field when a plane wave excites the cylinder as well as in both
the scattering amplitude and the total scattering cross section.
As far as the scattering by a plane wave propagating along the $x$
axis is concerned, the total magnetic field in region I is given
by
\begin{eqnarray}\label{Ctotal}
&  & H_z^{\mathrm I}(\rho ,\theta) = e^{ik^{\mathrm {I}}\rho \cos
\theta} \nonumber \\
&  & \qquad + \sum_{\ell=0}^{+\infty} \frac{\gamma_\ell}{2} \,
i^\ell \, \left[S_\ell - 1 \right] H_\ell^{(1)} (k^{\mathrm
{I}}\rho)\cos ({\ell\theta}).
\end{eqnarray}
Here $\gamma_\ell$ is the Neumann factor ($\gamma_0=1$ and for
$\ell \not= 0$,  $\gamma_\ell=2$).  The scattering amplitude
$f(\omega, \theta)$ is defined from the asymptotic behavior of the
total magnetic field by
\begin{equation}\label{defSA}
H_z^{\mathrm I}(\rho ,\theta) \underset{\rho \to +\infty}{\sim}
e^{ik^{\mathrm {I}}\rho \cos \theta} + f(\omega, \theta)
\frac{e^{i k^{\mathrm {I}}\rho}}{\sqrt{\rho}}.
\end{equation}
By using the asymptotic behavior (\ref{AsympHank1}) in
Eq.~(\ref{Ctotal}), we can write
\begin{equation}\label{ampli}
f(\omega, \theta)=\sqrt{\frac{1}{2i\pi k^{\mathrm {I}}(\omega)}}
\sum_{\ell=0}^{+\infty} \gamma_\ell \left[S_\ell(\omega) - 1
\right] \cos ({\ell\theta}).
\end{equation}
Then, the total scattering cross section per unit length of the
cylinder can be obtained by using the optical theorem
\cite{New82}:
\begin{equation}\label{CS}
\sigma _T(\omega)=\sqrt{\frac{8\pi}{k^{\mathrm {I}}(\omega)}}
\mathrm{Im} \left( e^{-i\pi/4}f(\omega, \theta =0)\right).
\end{equation}

\begin{figure}
\includegraphics[height=7cm,width=8.6cm]{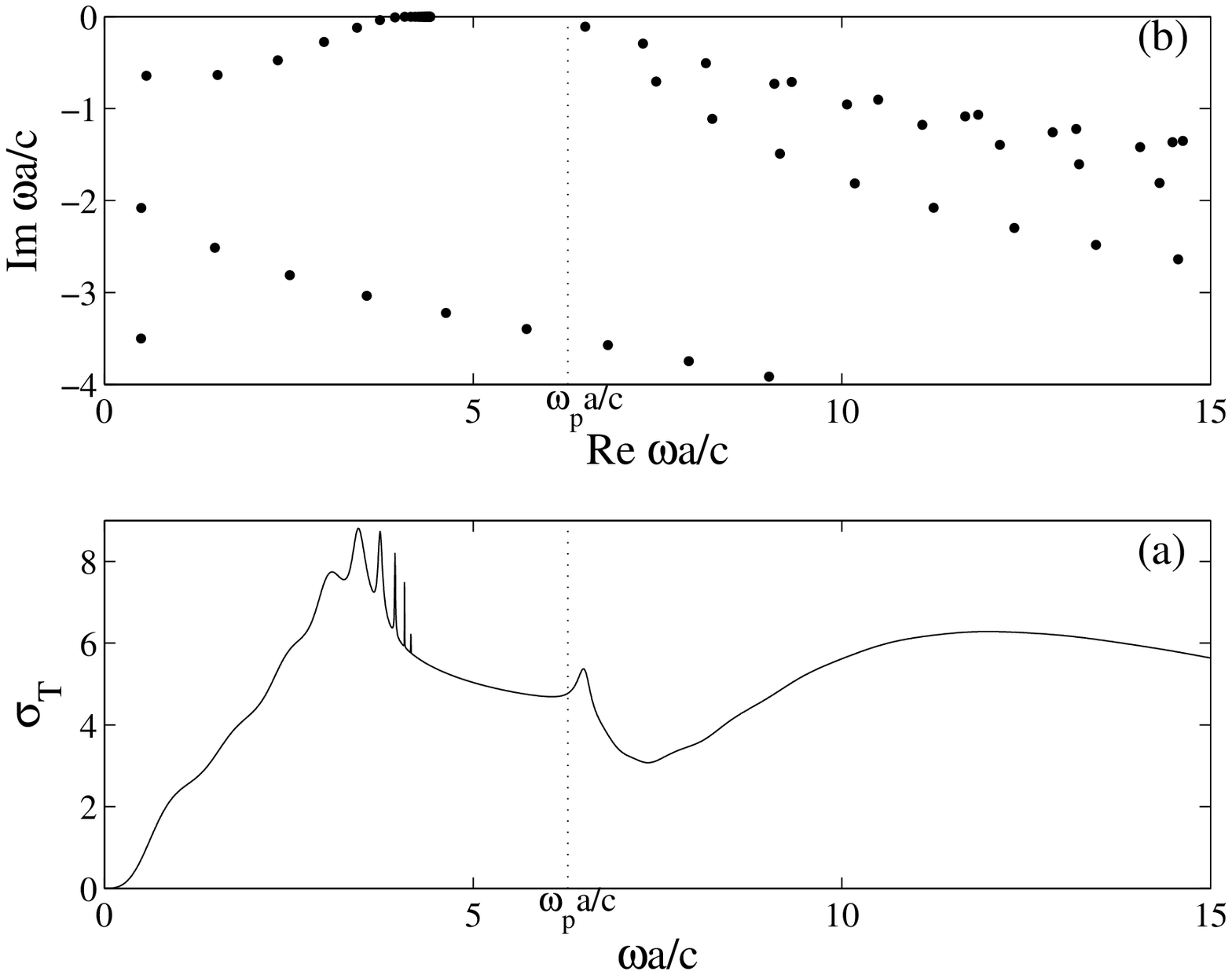}
\caption{\label{fig:crossMet} a) Total cross section $\sigma _T$.
b) Scattering resonances in the complex $\omega a/c$-plane.
$\epsilon_c(\omega)$ has the Drude type behavior with
$\epsilon_\infty=1$ and $\omega_pa/c=2\pi$ while $\epsilon_h=1$.}
\end{figure}
\begin{figure}
\includegraphics[height=7cm,width=8.6cm]{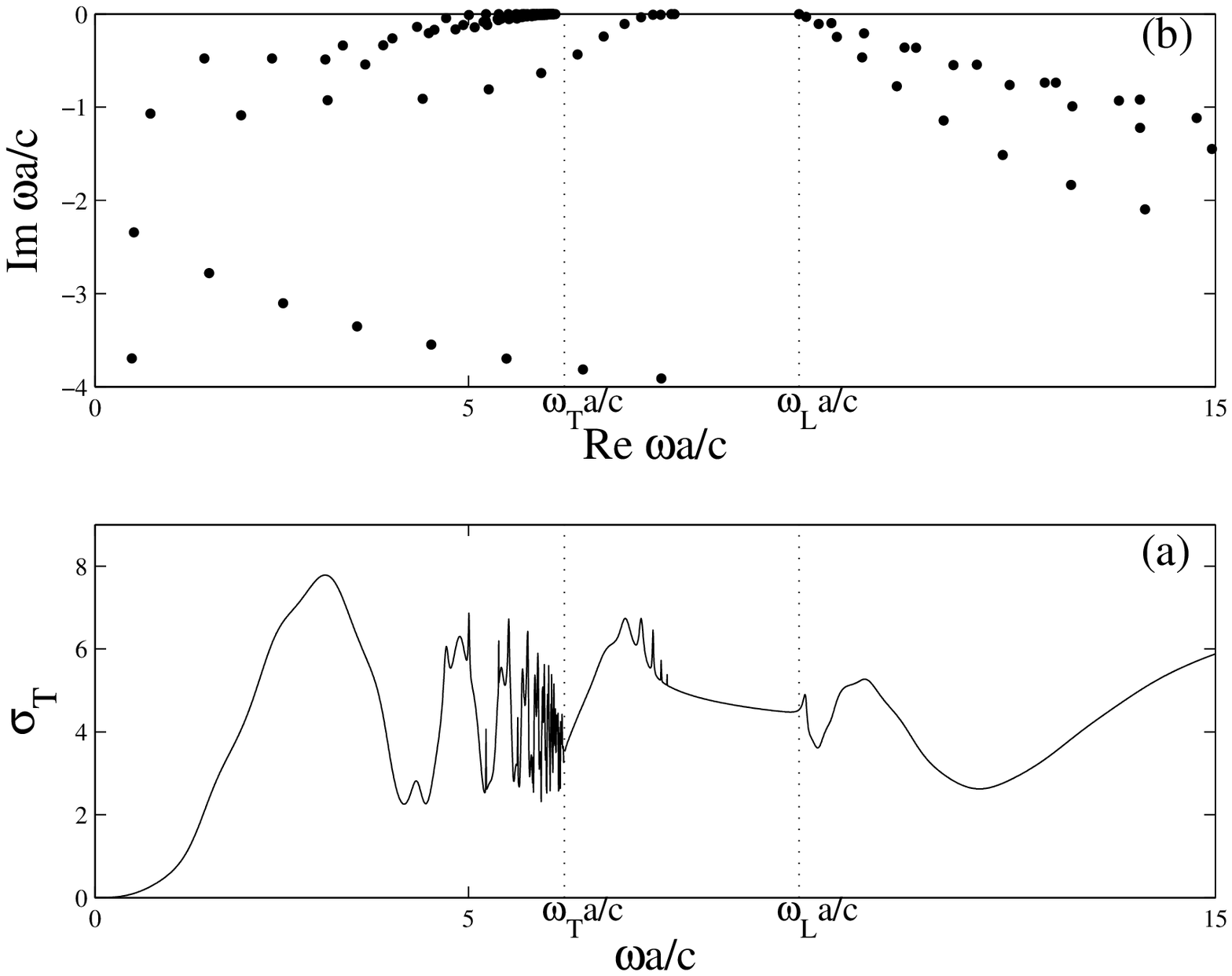}
\caption{\label{fig:crossSC} a) Total cross section $\sigma _T$.
b) Scattering resonances in the complex  $\omega a/c$-plane.
$\epsilon_c(\omega)$ has the ionic crystal behavior with
$\epsilon_\infty=1$, $\omega_Ta/c=2\pi$ and $\omega_La/c=3\pi$
while $\epsilon_h=1$.}
\end{figure}

In Figs.~\ref{fig:crossMet}a and \ref{fig:crossSC}a, we present
two examples of total cross section. They are both plotted as
functions of the reduced frequency $\omega a /c$. In
Fig.~\ref{fig:crossMet}a, the cylinder is embedded in vacuum
($\epsilon_h=1$) and its dielectric function is given by
(\ref{PermDrude}) with $\epsilon_\infty=1$ and $\omega_pa/c=2\pi$.
In Fig.~\ref{fig:crossSC}a, the cylinder is embedded in vacuum
($\epsilon_h=1$) and its dielectric function is given by
(\ref{PermCristIon}) with $\epsilon_\infty=1$, $\omega_Ta/c=2\pi$
and $\omega_La/c=3\pi$. Even if we restrict ourselves to those
particular configurations, it should be noted that the results
emphasized numerically and that we shall now discuss are very
general. On the two figures, rapid variations of sharp
characteristic shapes can be observed. This strongly fluctuating
behavior is due to scattering resonances. These resonances are the
poles of the $S$-matrix lying in the fourth quadrant of the
complex $\omega$-plane and they are determined by solving
\begin{equation}\label{det}
D_\ell(\omega)=0 \quad \mathrm{for} \quad \ell \in \mathbf{N}.
\end{equation}
The solutions of (\ref{det}) are denoted by $\omega_{\ell
p}=\omega^{(o)}_{\ell p}-i\Gamma _{\ell p}/2$ where
$\omega^{(o)}_{\ell p}>0$ and $\Gamma _{\ell p}>0$, the index $p$
permitting us to distinguish between the different roots of
(\ref{det}) for a given $\ell$. In the immediate neighborhood of
the resonance $\omega_{\ell p}$, $S_\ell(\omega)$ has the
Breit-Wigner form, i.e., is proportional to
\begin{equation}\label{BW}
\frac{\Gamma _{\ell p}/2}{\omega -\omega^{(o)}_{\ell p}+i\Gamma
_{\ell p}/2}.
\end{equation}
As a consequence, when a pole of the $S$-matrix is sufficiently
close to the real axis in the complex $\omega$-plane, it has an
appreciable influence on the scattering amplitude and therefore on
the total cross section. Of course, if a pole is very close to
this axis, the corresponding peak is too sharp to be observed on
the total cross section. In Figs.~\ref{fig:crossMet}b and
\ref{fig:crossSC}b, resonances are exhibited for the two
configurations previously considered. A one-to-one correspondence
between the peaks of $\sigma _T$ and the resonances near the real
$\omega a/c$-axis can be clearly observed in certain frequency
ranges.

More precisely and more generally, for the dielectric function
(\ref{PermDrude}) there exists in the frequency range $\omega <
\omega_p$ where $\epsilon_c (\omega) <0$ a family of resonances
close to the real axis of the complex $\omega$-plane which
converges, for large $\ell$, to the limiting frequency $\omega_s$
satisfying
\begin{equation}\label{accFREQmsc}
\epsilon_c(\omega_s) + \epsilon_h =0
\end{equation}
and given by
\begin{equation}\label{accFREQm1}
\omega_s =  \frac{\omega_p}{ \sqrt{1+ \epsilon_h /
\epsilon_\infty}}.
\end{equation}
For the dielectric function (\ref{PermCristIon}), the same type of
behavior occurs but in two different frequency ranges: in the
region $\omega < \omega_T$ (here $\epsilon_c (\omega) > 0$) with
an accumulation of resonances at the pole $\omega_T$ of the
dielectric function and in the region $\omega_T < \omega <
\omega_L$ (here $\epsilon_c (\omega) < 0$) with an accumulation of
resonances at the limiting frequency $\omega_s$ still satisfying
(\ref{accFREQmsc}) but which is now given by
\begin{equation}\label{accFREQsc2}
\omega_s = \sqrt{\frac{\omega_L^2 + (\epsilon_h / \epsilon_\infty)
\omega_T^2 }{ 1+ \epsilon_h / \epsilon_\infty }}.
\end{equation}
We must keep in mind that in the scattering of a $H$-polarized
photon with frequency $\omega^{(o)}_{\ell p}$, a decaying state of
the photon-cylinder system is formed. It has a finite lifetime
proportional to $1/\Gamma _{\ell p}$. The resonant states whose
complex frequencies belong to one of the families previously
described are therefore long-lived states. Because of these
particular quasibound states, the photon-cylinder system behaves
as a kind of artificial atom for which the photon plays the usual
role of the electron. We shall come back to this point of view in
the next two sections.

From now on, we shall more particularly focus our attention on the
physical interpretation of the long-lived resonant states whose
excitation frequencies belong to frequency ranges in which
$\epsilon_c (\omega) <0$. We shall prove that these states are
generated by a SP propagating close to the cylinder surface and
for this reason we call them RSPM's. For such states, the
artificial-atom point of view can be pushed farther as we shall
show in Sections 3 and 4. For the long-lived resonant states whose
excitation frequencies belong to the frequency range in which
$\epsilon_c (\omega) > 0$ (the so-called bulk polariton states),
we are not able to provide a similar analysis. This is not very
serious as they do not have, in photonic crystal physics, the
importance of RSPM's.

\section{Semiclassical analysis: From the SP Regge pole to
the complex frequencies of RSPM's}

\begin{figure}
\includegraphics[height=5.0cm,width=8.6cm]{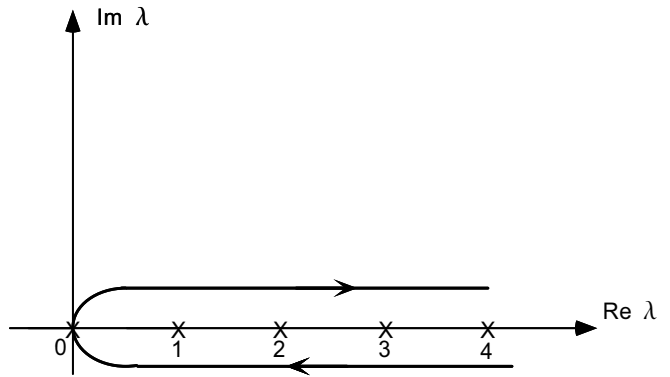}
\caption{\label{fig:watson} The Watson integration contour. }
\end{figure}

Using the CAM method, we can provide a physical picture of the
scattering process in term of diffraction by surface waves and
more particularly a physical explanation of the excitation
mechanism of RSPM's valid for ``high frequencies". By means of a
Watson transformation \cite{Watson18} applied to the scattering
amplitude (\ref{ampli}), we can write
\begin{equation}\label{ampliII}
f(\omega, \theta)=-\sqrt{ \frac{i} {2\pi k^{\mathrm {I}}(\omega)}}
~ {\mathcal P} \int_{\cal C} \frac{\left(S_{\lambda} (\omega) - 1
\right)}{\sin \pi \lambda} \cos \left[\lambda (\pi -\theta)\right]
d\lambda .
\end{equation}
Here $\mathcal{C}$ is the integration contour in the complex
$\lambda$-plane \cite{Watson18} illustrated in
Fig.~\ref{fig:watson} and which encircles the real axis in the
clockwise sense. In Eq.~(\ref{ampliII}), ${\mathcal P}$ which
stands for Cauchy's principal value at the origin is used in order
to reproduce the Neumann factor.  The Watson transformation has
permitted us to replace the ordinary angular momentum $\ell $ by
the complex angular momentum $\lambda$. $S_{\lambda} (\omega)$ is
now an analytic extension of $S_\ell (\omega )$ into the complex
$\lambda$-plane which is regular in the vicinity of the positive
real $\lambda$ axis. Using Cauchy's theorem and by noting that
inside the contour $\mathcal{C}$, the only singularities of the
integrand in (\ref{ampliII}) are the integers, we can easily
recover (\ref{ampli}) from (\ref{ampliII}).

\begin{figure}
\includegraphics[height=6cm,width=8.6cm]{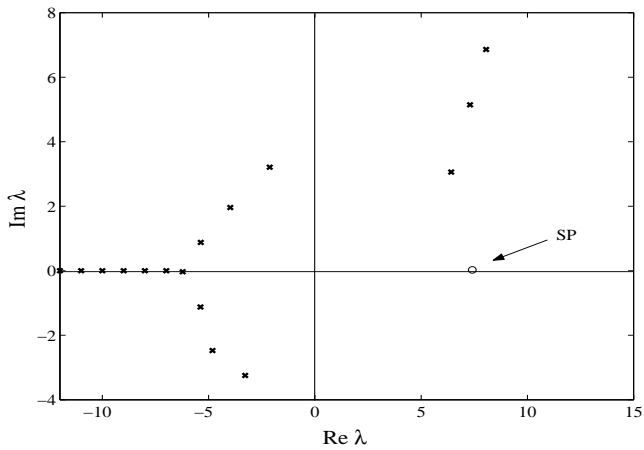}
\caption{\label{fig:RP1} Regge poles in the complex angular
momentum plane. $\epsilon_c(\omega)$ has the Drude type behavior
with $\epsilon_\infty=1$ and $\omega_pa/c=2\pi$ while
$\epsilon_h=1$. The distribution corresponds to $\omega a/c= 4$.}
\end{figure}

\begin{figure}
\includegraphics[height=6cm,width=8.6cm]{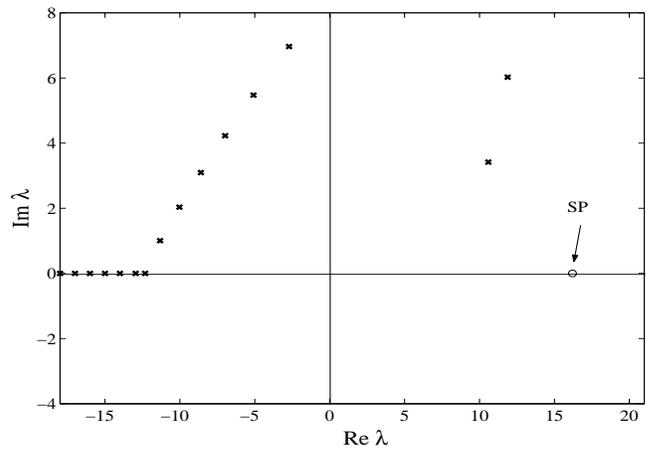}
\caption{\label{fig:RP2} Regge poles in the complex angular
momentum plane. $\epsilon_c(\omega)$ has the ionic crystal
behavior with $\epsilon_\infty=1$, $\omega_Ta/c=2\pi$ and
$\omega_La/c=3\pi$ while $\epsilon_h=1$. The distribution
corresponds to $\omega a/c=7.8$.}
\end{figure}

We can then deform the path of integration in (\ref{ampliII})
taking into account the possible singularities. The only
singularities that are encountered are the poles of the S-matrix
lying in the CAM plane. They are known as Regge poles
\cite{New82,Nus92} and are determined by solving
\begin{equation}\label{RP}
D_\lambda( \omega)=0 \quad \mathrm{for} \quad \omega > 0.
\end{equation}
Figs.~\ref{fig:RP1} and \ref{fig:RP2} exhibit the distribution of
Regge poles for a cylinder embedded in vacuum. We still consider
the two configurations previously studied. We do not display the
Regge pole distributions for other configurations (i.e. for other
values of the parameters $\epsilon_\infty$, $\epsilon_h$,
$\omega_P$, $\omega_T$ or $\omega_L$) because they are not really
different from those of Figs.~\ref{fig:RP1} and \ref{fig:RP2}. In
fact, all these Regge pole distributions are rather similar to the
distributions associated with the dielectric objects usually
studied \cite{Nus92}. However, in the frequency range where
$\epsilon (\omega) <0$, something new occurs: it exists a
particular Regge pole lying in the first quadrant of the
$\lambda$-plane and very close to the real axis. It is not present
for ordinary dielectric objects. As we shall see below, this new
Regge pole is associated with a SP orbiting around metallic or
semiconducting cylinders. From now on, we shall denote it by
$\lambda_\mathrm{SP}(\omega)$.

By Cauchy's theorem we can then extract from (\ref{ampliII}) the
contribution of a residue series over Regge poles.  In fact, we
limit our study to the contribution of
$\lambda_\mathrm{SP}(\omega)$ which is given by
\begin{equation}\label{ampliIIIa}
f_\mathrm{SP}(\omega, \theta)=\sqrt{\frac{2\pi }{i k^{\mathrm
{I}}(\omega)}} \frac{r_\mathrm{SP}(\omega)}{\sin \left[\pi
\lambda_\mathrm{SP}(\omega)\right]} \cos
\left[\lambda_\mathrm{SP}(\omega) (\pi -\theta)\right].
\end{equation}
Here $r_\mathrm{SP}(\omega)=\mathrm{residue}
\left(S_\lambda(\omega)\right)_{\lambda = \lambda
_\mathrm{SP}(\omega)}$.  Of course, $f$ differs from
$f_\mathrm{SP}$ by a smooth background integral and by the
contributions of all other Regge poles. We are not interested by
these contributions which do not play any role in the excitation
of RSPM's. We think that these contributions could be studied,
{\it mutatis mutandis}, in the framework of CAM techniques
developed by Nussenzveig and coworkers \cite{Nus92}. By using
\begin{equation}
\frac{1}{\sin \pi \lambda }=-2i \sum_{m=0}^{+\infty} e^{
i\pi(2m+1)\lambda} \nonumber
\end{equation}
which is true if $\mathrm{Im} \ \lambda > 0$, we can write
\begin{eqnarray}\label{ampliIIIb}
&   &   f_\mathrm{SP}(\omega, \theta)=-\sqrt{\frac{2i\pi
}{k^{\mathrm {I}}(\omega)}}~
r_\mathrm{SP}(\omega) \nonumber \\
&  & \qquad  \qquad  \times \sum_{m=0}^{+\infty}  \left(
e^{i\lambda_\mathrm{SP} (\omega
)(\theta +2m\pi) } \right. +  \nonumber \\
&  & \qquad \qquad \qquad \qquad  \left.
e^{i\lambda_\mathrm{SP}(\omega )(2\pi - \theta +2m\pi)} \right).
\quad
\end{eqnarray}
In Eq.~(\ref{ampliIIIb}), exponential terms correspond to surface
wave contributions. Because the Regge pole
$\lambda_\mathrm{SP}(\omega )$ lies in the first quadrant of the
CAM plane, $\exp[i\lambda_\mathrm{SP}(\omega )(\theta)]$ (resp.
$\exp[i\lambda_\mathrm{SP}(\omega )(2\pi - \theta)]$) corresponds
to a surface wave propagating counterclockwise (resp. clockwise)
around the cylinder and $\mathrm{Re} \
\lambda_\mathrm{SP}(\omega)$ represents its azimuthal propagation
constant while $\mathrm{Im} \ \lambda_\mathrm{SP}(\omega)$ is its
damping constant. The exponential decay $\exp[-\mathrm{Im} \
\lambda_\mathrm{SP}(\omega)\theta]$ (resp. $\exp[-\mathrm{Im} \
\lambda_\mathrm{SP}(\omega)(2\pi - \theta)]$) is due to continual
re-radiation of energy. Moreover, in (\ref{ampliIIIb}), the sum
over $m$ takes into account the multiple circumnavigations of the
surface waves around the cylinder as well as the associated
radiation damping. The Regge pole $\lambda_\mathrm{SP}$ is very
close to the real axis in the complex $\lambda$-plane. It then
corresponds to a surface wave which is slightly attenuated during
its propagation and which contributes significantly to the
scattering process and to the resonance mechanism.


\begin{figure}
\includegraphics[height=6cm,width=8.6cm]{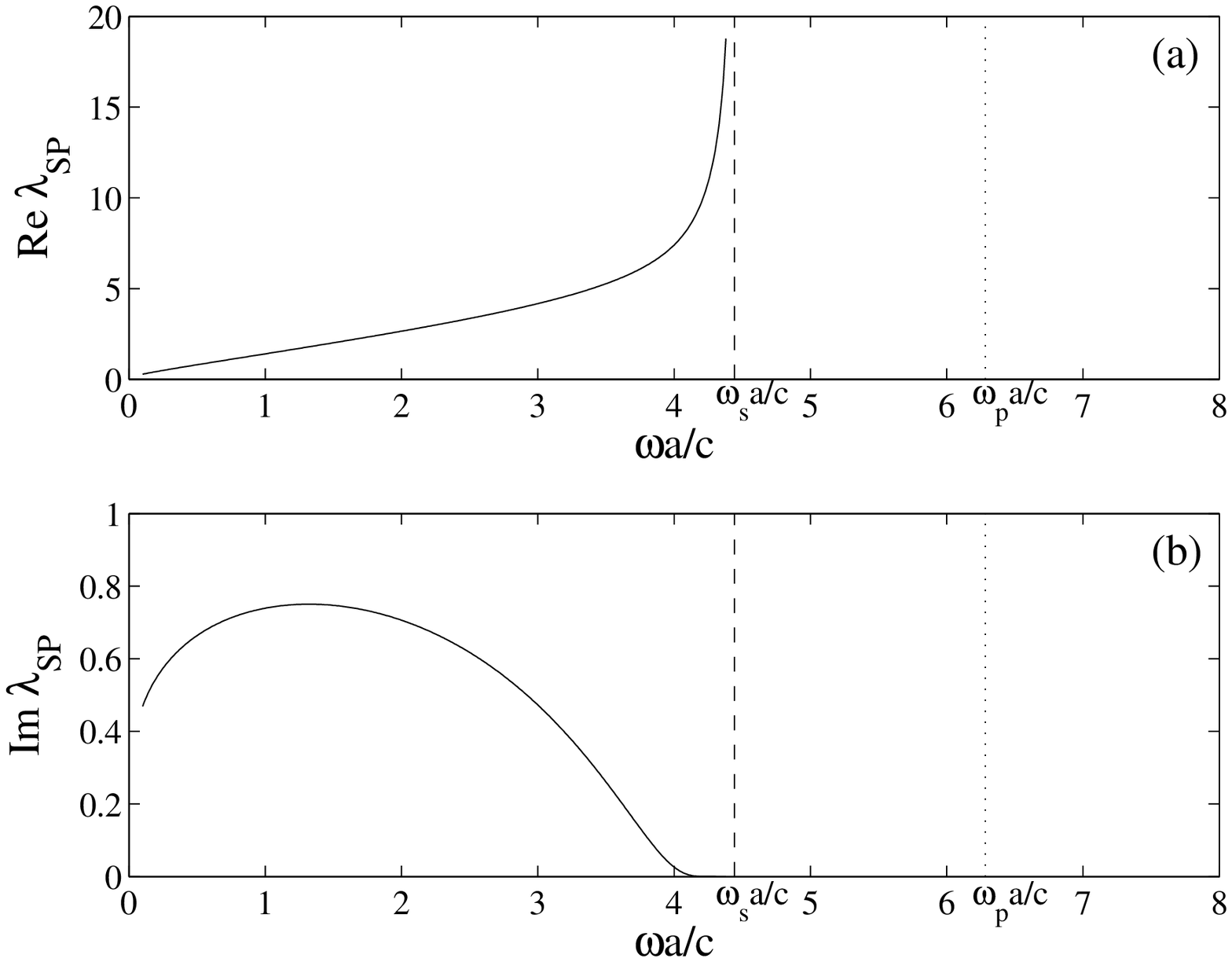}
\caption{\label{fig:RT1} Regge trajectory for the SP Regge pole.
$\epsilon_c(\omega)$ has the Drude type behavior with
$\epsilon_\infty=1$ and $\omega_pa/c=2\pi$ while $\epsilon_h=1$.
As $\omega a/c \to \omega_s a/c$, the real part of the SP Regge
pole increases indefinitely  while its imaginary part vanishes.}
\end{figure}

\begin{figure}
\includegraphics[height=6cm,width=8.6cm]{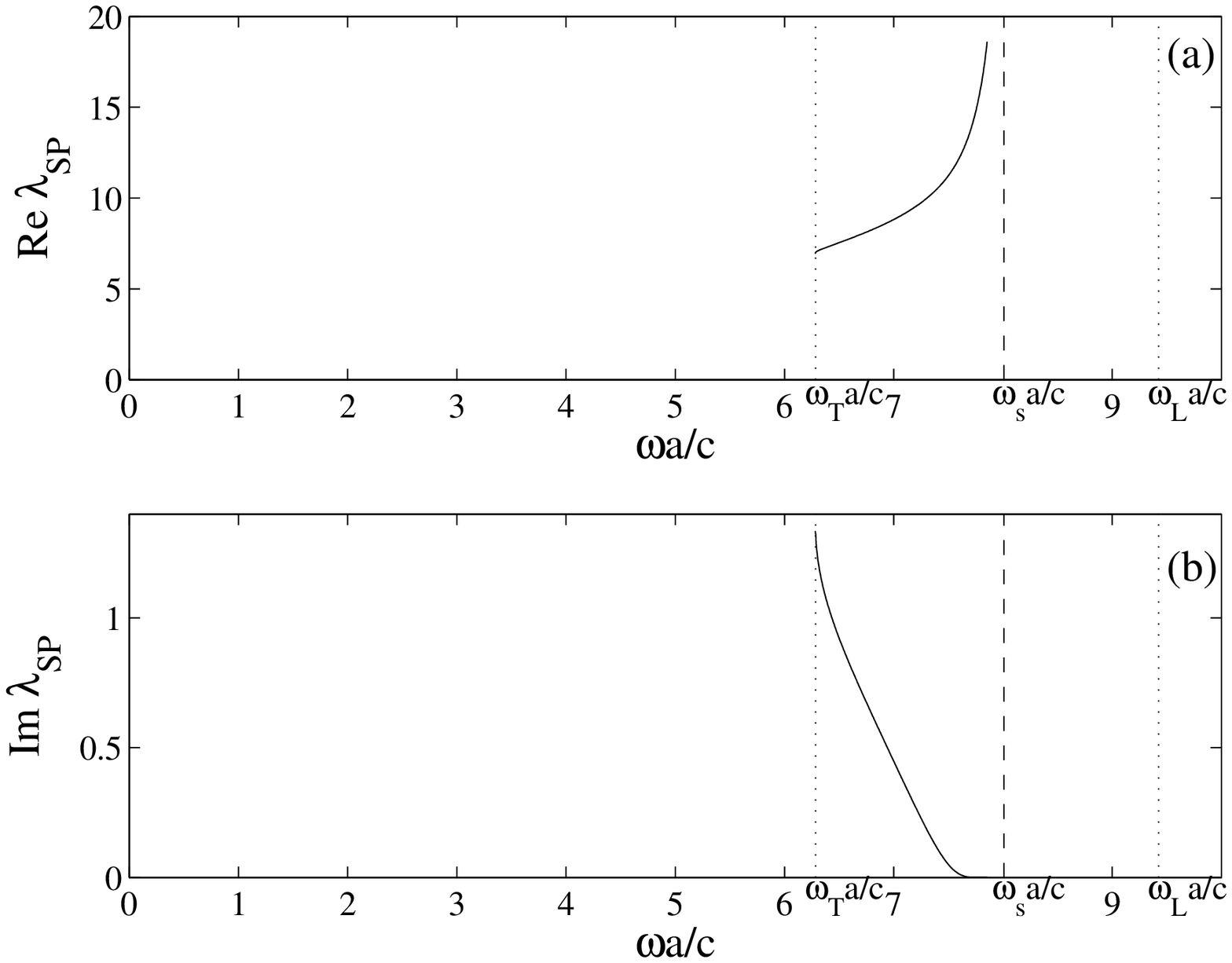}
\caption{\label{fig:RT2}Regge trajectory for the SP Regge pole.
$\epsilon_c(\omega)$ has the ionic crystal behavior with
$\epsilon_\infty=1$, $\omega_Ta/c=2\pi$ and $\omega_La/c=3\pi$
while $\epsilon_h=1$. As $\omega a/c \to \omega_s a/c$, the real
part of the SP Regge pole increases indefinitely while its
imaginary part vanishes.}
\end{figure}

As $\omega $ varies, the Regge pole $\lambda_\mathrm{SP}(\omega )$
describes a Regge trajectory \cite{New82} in the CAM plane. In
Figs.~\ref{fig:RT1} and \ref{fig:RT2}, we have displayed the Regge
trajectories of SP's for the two configurations previously
studied. It should be noted that as $\omega \to \omega_s $, the
real part of the SP Regge pole increases indefinitely while its
imaginary part vanishes. For other configurations (i.e. for other
values of the parameters $\epsilon_\infty$, $\epsilon_h$,
$\omega_p$, $\omega_T$ or $\omega_L$), we have verified that the
SP Regge pole behavior is very similar.

The resonant behavior of the cylinder-photon system can now be
understood in terms of the Regge trajectory of the SP. When the
quantity $\mathrm{Re} \ \lambda_\mathrm{SP}(\omega )$ coincides
with an integer, a resonance occurs. Indeed, it is produced by a
constructive interference between the different components of the
surface wave, each component corresponding to a different number
of circumnavigations. Resonance excitation frequencies
$\omega^{(o)}_{\ell \mathrm{SP}}$ are therefore obtained from the
Bohr-Sommerfeld type quantization condition
\begin{equation}\label{sc1}
\mathrm{Re} \ \lambda_\mathrm{SP} \left(\omega^{(o)}_{\ell
\mathrm{SP}} \right)= \ell  \qquad \ell =0,1,2,\dots .
\end{equation}
By assuming that $\omega $ is in the neighborhood of
$\omega^{(o)}_{\ell \mathrm{SP}}$ and using $\mathrm{Re} \
\lambda_\mathrm{SP} (\omega ) \gg \mathrm{Im} \
\lambda_\mathrm{SP} (\omega )$ (which can be numerically verified,
except for low frequencies), we can expand $\lambda_\mathrm{SP}
(\omega)$ in a Taylor series about $\omega^{(o)}_{\ell ~
\mathrm{SP}}$, and obtain
\begin{eqnarray} \label{TS}
& & \lambda_\mathrm{SP}(\omega) \approx \ell  +  \left. \frac{d
\mathrm{Re} \lambda_\mathrm{SP}(\omega)}{d\omega} \right|_{\omega
=\omega^{(o)}_{\ell \mathrm{SP}}} (\omega - \omega^{(o)}_{\ell
\mathrm{SP}}  ) \nonumber \\
& & \qquad \qquad \quad + \, i~ \mathrm{Im} \lambda_\mathrm{SP}
(\omega^{(o)}_{\ell \mathrm{SP}}) + \dots .
\end{eqnarray}
Then, by replacing (\ref{TS}) in the term ${\cos \left[\pi
\lambda_\mathrm{SP}(\omega )  \right]}$ of (\ref{ampliIIIa}), we
show that $f_\mathrm{SP}(\omega, \theta)$ presents a resonant
behavior given by the Breit-Wigner formula (\ref{BW}) with
\begin{equation}\label{sc2}
\frac{\Gamma _{\ell \mathrm{SP}}}{2}= \left.  \frac{\mathrm{Im} \
\lambda_\mathrm{SP} (\omega )}{d \ \mathrm{Re} \
\lambda_\mathrm{SP} (\omega ) /d\omega } \right|_{\omega
=\omega^{(o)}_{\ell \mathrm{SP}}}.
\end{equation}
Eqs.~(\ref{sc1}) and (\ref{sc2}) are semiclassical formulas which
permit us to determine the location of the resonances from the
Regge trajectory of $\lambda_\mathrm{SP}$.


Tables~\ref{tab:table1} and \ref{tab:table2} present samples of
complex frequencies of RSPM's for the two configurations
previously considered. They are calculated from the semiclassical
formulas (\ref{sc1}) and (\ref{sc2}) by using the Regge
trajectories numerically determined by solving (\ref{RP}) (see
Figs.~\ref{fig:RT1} and \ref{fig:RT2}). A comparison between the
``exact" and the semiclassical spectra shows a very good
agreement, except for ``low" frequencies. We have also performed
the corresponding calculations for other configurations with
$\epsilon_h \not=1$ and $\epsilon_\infty \not=1$. The agreement
seems even better. Furthermore, inserted into the semiclassical
formulas (\ref{sc1}) and (\ref{sc2}), the behavior of Regge
trajectories near the limiting frequencies $\omega_s$ easily
explains the existence of the families of resonances close to the
real axis of the complex $\omega$-plane which converges for large
$\ell$ to the limiting frequency $\omega_s$. In conclusion, we
have established a connection between the complex frequencies of
RSPM's and a particular surface wave, the so-called SP, described
by a particular Regge pole of the $S$-matrix and which orbits
around the cylinder.

\begin{table}
\caption{\label{tab:table1} The first complex frequencies of
RSPM's. $\epsilon_c(\omega)$ has the Drude type behavior with
$\epsilon_\infty=1$ and $\omega_p a/c=2\pi$ while $\epsilon_h=1$.}
\begin{ruledtabular}
\begin{tabular}{ccccc}
        &\quad Exact \quad& \quad
Exact \quad&  Semiclassical  &  Semiclassical  \\
 $\ell $   &$   \omega ^{(o)}_{\ell {\mathrm SP}}  $
 &  $  \Gamma _{\ell {\mathrm SP}}/2  $
 & $  \omega ^{(o)}_{\ell {\mathrm SP}}   $ &  $ \Gamma _{\ell {\mathrm SP}}/2  $  \\
\hline
1&    0.570278   &  0.642122   & 0.665828  & 0.582435 \\
2&   1.53524   & 0.633630   &  1.49245  & 0.602114  \\
3&    2.35038   &  0.475286  &  2.25693  & 0.476399    \\
4&   2.97766    & 0.274602   &  2.90439  & 0.289791   \\
5&  3.42645    &   0.119226  &  3.39638  & 0.127238    \\
6&   3.73632   & 0.036765   &  3.73016  & 0.037645   \\
7&  3.94064     & 0.007494   & 3.94009  &  0.007435  \\
8&  4.07049    &  0.000999  & 4.07047   & 0.000981   \\
9&   4.15483    &  0.000093  & 4.15476   & 0.000094  \\
10&   4.21272    &  0.000006  &  4.21266 & 0.000006
\end{tabular}
\end{ruledtabular}
\end{table}
\begin{table}
\caption{\label{tab:table2} The first complex frequencies of
RSPM's. $\epsilon_c(\omega)$ has the ionic crystal behavior with
$\epsilon_\infty=1$, $\omega_Ta/c=2\pi$ and $\omega_La/c=3\pi$
while $\epsilon_h=1$.}
\begin{ruledtabular}
\begin{tabular}{ccccc}
        &\quad Exact \quad& \quad
Exact \quad&  Semiclassical  &  Semiclassical  \\
 $\ell $   &$   \omega ^{(o)}_{\ell {\mathrm SP}}  $
 &  $  \Gamma _{\ell {\mathrm SP}}/2  $
 & $  \omega ^{(o)}_{\ell {\mathrm SP}}   $ &  $ \Gamma _{\ell {\mathrm SP}}/2  $  \\
\hline
7&   6.459387    &  0.4345112   & 6.283821     & 0.0837525  \\
8&   6.809183   &  0.2419456   & 6.698432    &  0.3004512  \\
9&   7.091176    & 0.1055079   & 7.056186    &  0.1188560  \\
10&  7.311233  &   0.0353546  & 7.304271     & 0.0371596   \\
11&  7.470835    &   0.0088275  & 7.470046     & 0.0088537   \\
12&  7.581710     &  0.0016441  & 7.581652   &   0.0016417  \\
13&  7.659584    &    0.0002366  & 7.659579    &  0.0002392  \\
14&  7.716557    & 0.0000273   & 7.716543    &  0.0000274  \\
15&  7.759919     & 0.0000026  &  7.759919   & 0.0000027
\end{tabular}
\end{ruledtabular}
\end{table}

\section{Semiclassical analysis: Asymptotics for
the SP and physical description}

\begin{figure}
\includegraphics[height=6cm,width=8.6cm]{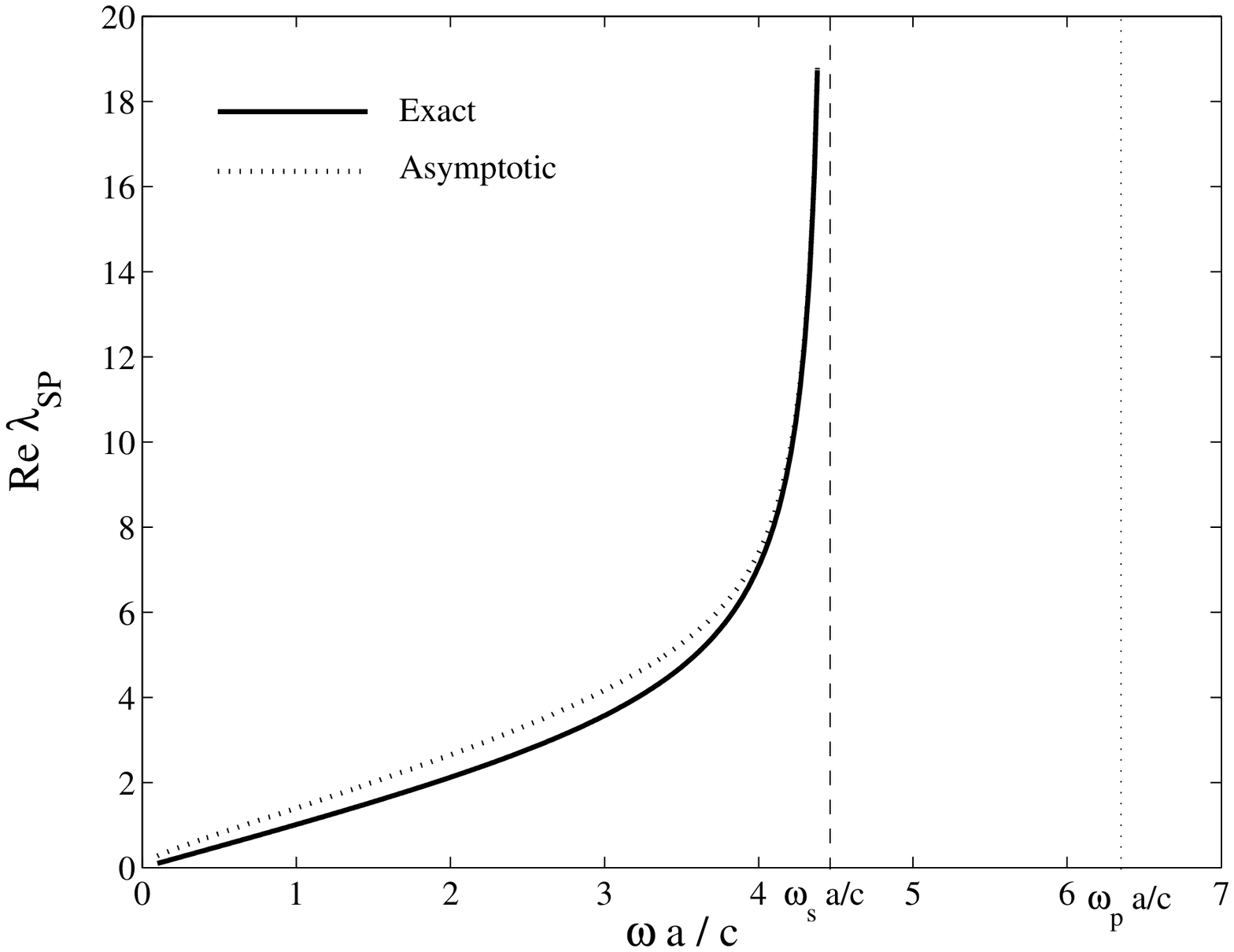}
\caption{\label{fig:RT1asymp} Regge trajectory for the SP Regge
pole. Comparison between exact and asymptotic theories.
$\epsilon_c(\omega)$ has the Drude type behavior with
$\epsilon_\infty=1$ and $\omega_pa/c=2\pi$ while $\epsilon_h=1$.}
\end{figure}

\begin{figure}
\includegraphics[height=6cm,width=8.6cm]{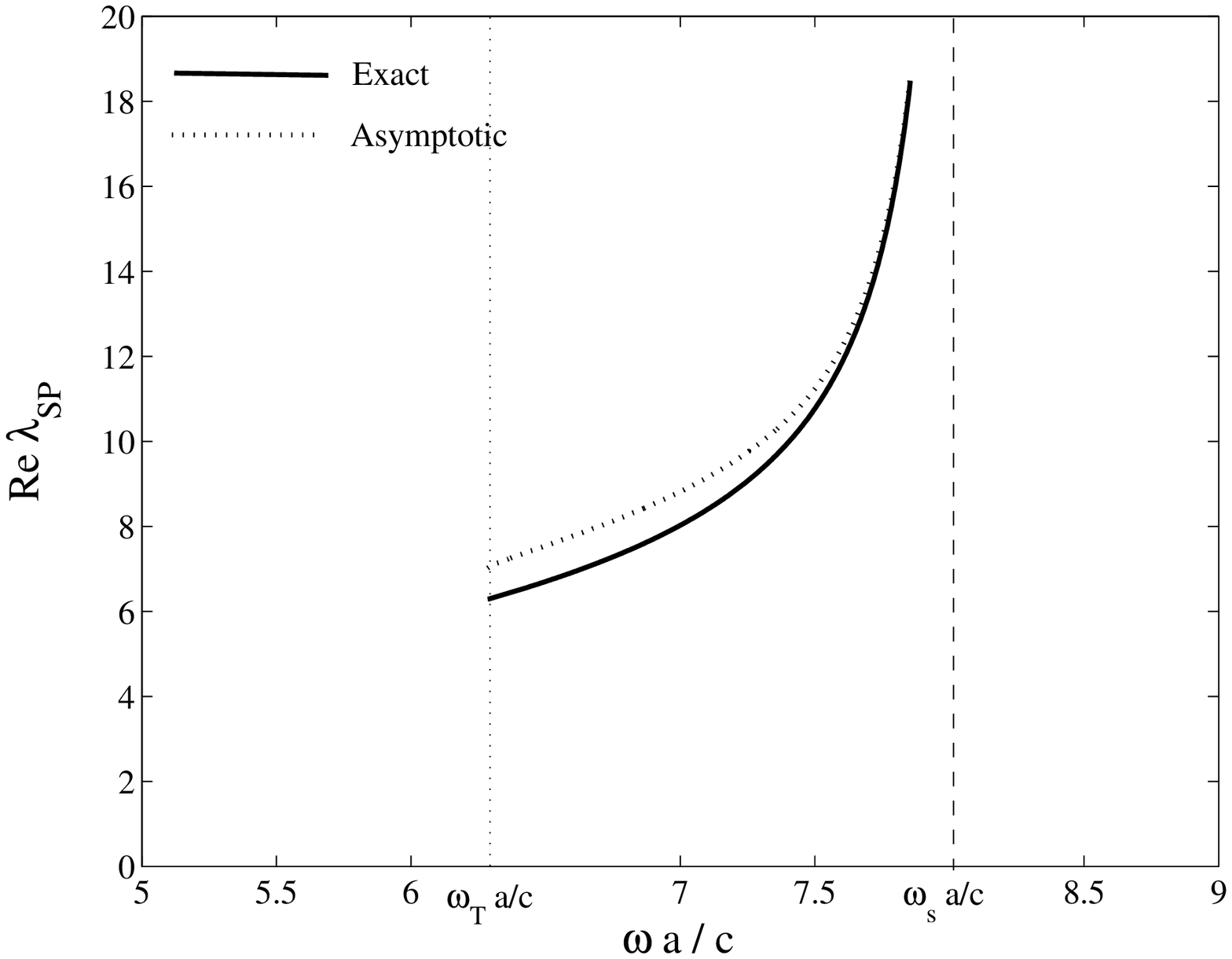}
\caption{\label{fig:RT2asymp}Regge trajectory for the SP Regge
pole. Comparison between exact and asymptotic theories.
$\epsilon_c(\omega)$ has the ionic crystal behavior with
$\epsilon_\infty=1$, $\omega_Ta/c=2\pi$ and $\omega_La/c=3\pi$
while $\epsilon_h=1$.}
\end{figure}

\begin{figure}
\includegraphics[height=6cm,width=8.6cm]{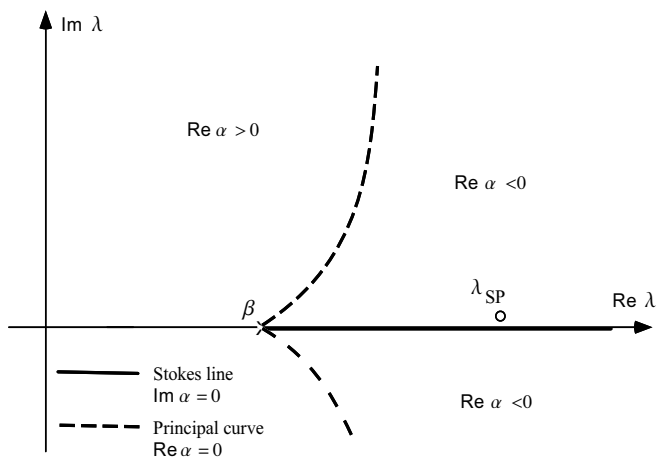}
\caption{\label{fig:StokesH1} The relative positions, in the
$\lambda$-complex plane, of the Regge pole $\lambda_\mathrm{SP}$
and the reduced frequency $\beta=\sqrt{\epsilon_h} \omega a/c$.}
\end{figure}

A deeper understanding of the SP behavior can be obtained by
solving perturbatively Eq.~(\ref{RP}) for $\lambda =
\lambda_\mathrm{SP}$. We first replace the Bessel function
$J_\lambda(z)$ by the modified Bessel function $I_\lambda(z)$ (see
Ref.~\onlinecite{AS65}) in order to take into account the fact
that $\epsilon_c(\omega) <0$. Eq.~(\ref{RP}) reduces to
\begin{equation} \label{RPSP1}
\frac{1}{\sqrt{\epsilon_h}}
\frac{H_{\lambda_\mathrm{SP}}^{(1)'}(\sqrt{\epsilon_h} a\omega/c
)}{H_{\lambda_\mathrm{SP}}^{(1)}(\sqrt{\epsilon_h} a\omega/c )} =
- \frac{1}{\sqrt{-\epsilon_c(\omega)}}
\frac{I'_{\lambda_\mathrm{SP}}(\sqrt{-\epsilon_c(\omega)}
a\omega/c )}{I_{\lambda_\mathrm{SP}}(\sqrt{-\epsilon_c(\omega)}
a\omega/c )}.
\end{equation}
In the right-hand side of (\ref{RPSP1}), we can use the uniform
asymptotic expansions of $I_\lambda(z)$ for large orders (see
Ref.~\onlinecite{AS65}), i.e.
\begin{equation}\label{AsympBesselI}
I_\lambda (z)  {\sim} \frac{1}{\sqrt{2\pi}} \frac{1}{(\lambda^2 +
z^2)^{1/4}} ~ e^{F_\lambda (z)/2}
\end{equation}
where
\begin{equation}
\frac{F_\lambda (z)}{2} = (\lambda ^2 + z^2)^{1/2} +  \lambda  \ln
\left( \frac{z}{\lambda + (\lambda ^2 + z^2)^{1/2}} \right).
\end{equation}
By assuming $|\lambda_\mathrm{SP}| \gg \sqrt{-\epsilon_c(\omega)}
a\omega/c$, we then obtain
\begin{eqnarray} \label{rhsRPSP1}
&  & - \frac{1}{\sqrt{-\epsilon_c(\omega)}}
\frac{I'_{\lambda_\mathrm{SP}}(\sqrt{-\epsilon_c(\omega)}
a\omega/c )}{I_{\lambda_\mathrm{SP}}(\sqrt{-\epsilon_c(\omega)}
a\omega/c )} \nonumber \\
&  & \qquad \qquad \qquad \sim \frac{\left[ \lambda_\mathrm{SP}^2
- \epsilon_c(\omega) (a\omega/c)^2
\right]^{1/2}}{\epsilon_c(\omega) (a\omega/c)}.
\end{eqnarray}
In the left-hand side of (\ref{RPSP1}), the relative positions of
$\lambda_\mathrm{SP}$ and $\sqrt{\epsilon_h} a\omega/c$ in the
$\lambda$-complex plane (see Fig.~\ref{fig:StokesH1}) permit us to
employ the Debye asymptotic expansion of $H_{\lambda}^{(1)}(z)$ in
the form \cite{WatsonBessel,Nuss65}
\begin{equation} \label{AsympDebyeIa}
H_{\lambda}^{(1)}(z) {\sim} -i A(\lambda ,z) e^{-\alpha(\lambda
,z)}
\end{equation}
where
\begin{subequations}
\begin{eqnarray}
&&A(\lambda ,z) = \left( \frac{2}{\pi} \right)^{1/2} (\lambda^2 - z^2)^{-1/4},
\label{AsympDebyeIb} \\
&& \alpha(\lambda ,z)   =  (\lambda^2 - z^2)^{1/2} -\lambda \ln
\left( \frac{z+ (\lambda^2 - z^2)^{1/2}}{z} \right).  \nonumber \\
&& \label{AsympDebyeIc}
\end{eqnarray}
\end{subequations}
By assuming $|\lambda_\mathrm{SP}| \gg \sqrt{\epsilon_h} a\omega/c
$, we then deduce
\begin{equation} \label{lhsRPSP1}
\frac{1}{\sqrt{\epsilon_h}}
\frac{H_{\lambda_\mathrm{SP}}^{(1)'}(\sqrt{\epsilon_h} a\omega/c
)}{H_{\lambda_\mathrm{SP}}^{(1)}(\sqrt{\epsilon_h} a\omega/c )}
\sim -  \frac{\left[ \lambda_\mathrm{SP}^2 - \epsilon_h
(a\omega/c)^2 \right]^{1/2}}{\epsilon_h (a\omega/c)}.
\end{equation}
Eq.~(\ref{RPSP1}) can now be solved and we easily find
\begin{equation} \label{ReRPSP}
\lambda_\mathrm{SP} (\omega ) \sim  \left( \frac{\omega a}{c}
\right) \sqrt{  \frac{\epsilon_h \epsilon_c(\omega )}{\epsilon_h +
\epsilon_c(\omega )}}.
\end{equation}
We have obtained an asymptotic expansion for $\lambda_\mathrm{SP}$
or more exactly for the real part of  $\lambda_\mathrm{SP}$.
Indeed, the right-hand side of (\ref{ReRPSP}) is purely real. The
perturbative resolution of Eq.~(\ref{RPSP1}) did not permit us to
extracted the small imaginary part of $\lambda_\mathrm{SP}$. Of
course, it would be possible to improve (\ref{ReRPSP}) by taking
into account higher orders in the asymptotic expansions
(\ref{AsympBesselI}) and (\ref{AsympDebyeIa}). But that does not
seem necessary. First, this does not provide an imaginary part for
$\lambda_\mathrm{SP}$. In fact, as we shall see below, this term
corresponds to an exponentially small contribution which lies
beyond all orders in perturbation theory. Furthermore, we have
numerically tested the formula (\ref{ReRPSP}) for various values
of the parameters $\epsilon_\infty$, $\epsilon_h$, $\omega_p$,
$\omega_T$ or $\omega_L$. In all cases, it provides a rather good
approximation for $\mathrm {Re} ~ \lambda_\mathrm{SP}$ (see, for
example, Figs.~\ref{fig:RT1asymp} and \ref{fig:RT2asymp} for the
two configurations previously studied). It should be noted that it
also predicts the divergence of $\mathrm {Re} ~
\lambda_\mathrm{SP}$ for $\omega \to \omega_s$.


We can now insert (\ref{ReRPSP}) into (\ref{ampliIIIb}). The
contributions corresponding to the SP propagating counterclockwise
and clockwise are then given by
\begin{equation} \label{ondesSP}
\exp \left[i( \pm \lambda_\mathrm{SP} (\omega ) \theta  - \omega
t) \right] = \exp \left[ i( \pm k_\mathrm{SP} (\omega ) a\theta
-\omega t)\right]
\end{equation}
with
\begin{equation} \label{DispRelSP}
k_\mathrm{SP}(\omega ) = \left( \frac{\omega}{c} \right) \sqrt{
\frac{\epsilon_h \epsilon_c(\omega )}{\epsilon_h +
\epsilon_c(\omega )}}.
\end{equation}
Here we have reintroduced the time dependence $\exp (-i\omega t)$
in order to clarify the physical interpretation. From
(\ref{ondesSP}) and by noting that $a \, d\theta$ represents the
length element on the cylinder surface, it now appears that the SP
propagation is supported by the cylinder surface which thus plays
the role of a Bohr-Sommerfeld-type orbit and that
(\ref{DispRelSP}) can be considered as the SP dispersion relation.
This relation could permit us to derive analytically the phase
velocity $v_p= \omega / k_\mathrm{SP}(\omega )$ as well as the
group velocity $v_g= d\omega / dk_\mathrm{SP}(\omega )$ of the SP.

It should be also noted that the dispersion relation
(\ref{DispRelSP}) is in fact the usual dispersion relation of a SP
supported by a flat metal-dielectric or semiconductor-dielectric
interface (see, for example Ref.~\onlinecite{Raether88}). We have
recovered the same dispersion relation because we have limited the
perturbative resolution of Eq.~(\ref{RPSP1}) to the lowest order.
By taking into account higher orders in the asymptotic expansions
(\ref{AsympBesselI}) and (\ref{AsympDebyeIa}), we could obtain
corrections for (\ref{ReRPSP}) and (\ref{DispRelSP}) vanishing for
$a \to \infty$, i.e. in the limit of large radius.

By inserting the expression (\ref{ReRPSP}) for
$\lambda_\mathrm{SP} (\omega )$ into the Bohr-Sommerfeld
quantization condition (\ref{sc1}), we can derive approximations
for the resonance excitation frequencies $\omega^{(o)}_{\ell
{\mathrm SP}}$. If the dielectric function of the cylinder is
given by (\ref{PermDrude}), we obtain for the reduced frequencies
\begin{widetext}
\begin{equation}
\frac{\omega^{(o)}_{\ell {\mathrm SP}}~a}{c}  \approx
\frac{1}{\sqrt{2}}
 \left[
\left(\frac{\omega_p a}{c}\right)^2 + \left( \frac{\epsilon_h+
\epsilon_\infty}{\epsilon_h \epsilon_\infty}\right) \ell^2 -\sqrt{
\left[ \left(\frac{\omega_p a}{c}\right)^2 + \left(
\frac{\epsilon_h+ \epsilon_\infty}{\epsilon_h
\epsilon_\infty}\right) \ell^2 \right]^2
-\frac{4}{\epsilon_h}\left(\frac{\omega_p a}{c}\right)^2 \ell^2 }
\,\, \right]^{1/2}. \label{ApproxREdrude}
\end{equation}
This analytic formula provides accurate results for ``large"
values of $\ell$. For $\ell =3$, the error is around $13~\%$ and
it becomes less than $1~\%$ for $\ell >7$. Furthermore, it
predicts the convergence to $\omega_s =  \omega_p / \sqrt{1+
\epsilon_h / \epsilon_\infty}$ when $\ell \to \infty$. For the
dielectric function (\ref{PermCristIon}), we obtain
\begin{eqnarray}
&  & \frac{\omega^{(o)}_{\ell {\mathrm SP}}~a}{c} \approx
\frac{1}{\sqrt{2}}
 \left[
\left(\frac{\omega_L a}{c}\right)^2 + \left( \frac{\epsilon_h+
\epsilon_\infty}{\epsilon_h \epsilon_\infty}\right) \ell^2  \quad
\phantom{\sqrt{ \left[ \left(\frac{\omega_L
a}{c}\right)^2\right]}} \right.
\nonumber \\
&  &  \qquad \left. - \sqrt{ \left[ \left(\frac{\omega_L
a}{c}\right)^2 + \left( \frac{\epsilon_h+
\epsilon_\infty}{\epsilon_h \epsilon_\infty}\right) \ell^2
\right]^2 -\frac{4}{\epsilon_h \epsilon_\infty }
\left[\epsilon_\infty \left(\frac{\omega_L a}{c}\right)^2 +
\epsilon_h \left(\frac{\omega_T a}{c}\right)^2\right] \ell^2 } \,
\right]^{1/2}. \label{ApproxREsc}
\end{eqnarray}
\end{widetext}
This analytic formula provides accurate results for ``large"
values of $\ell$. For $\ell =7$, the error is around $3~\%$ and it
becomes less than $1~\%$ for $\ell >10$. Furthermore, it predicts
the accumulation of resonances at $\omega_s = \sqrt{(\omega_L^2 +
(\epsilon_h / \epsilon_\infty) \omega_T^2)/(1+ \epsilon_h /
\epsilon_\infty)}$ for $\ell \to \infty$.

To conclude this section, we shall come back to the damping
constant $\mathrm{Im} \ \lambda_\mathrm{SP}(\omega)$ of the SP.
Numerically (see Figs.~\ref{fig:RT1} and \ref{fig:RT2}) we have
shown that this term is small but we cannot consider that it
vanishes as we previously found. In fact, this term corresponds to
an exponentially small contribution which lies beyond all orders
of the asymptotic expansion (\ref{AsympDebyeIa}) and which can be
captured by carefully taking into account Stokes' phenomenon
\cite{StokesPh1,StokesPh2}. (For modern aspects of asymptotics
beyond all orders and of Stokes' phenomenon, we refer to a
beautiful article by Berry \cite{Berry89} as well as to
Refs.~\onlinecite{Dingle73, SegurTL91,BerryHowls90}). Instead of
(\ref{AsympDebyeIa}), we shall use the Debye asymptotic expansion
of $H_{\lambda}^{(1)}(z)$ in the form
\begin{eqnarray} \label{AsympDebyeICORR}
&  & H_{\lambda}^{(1)}(z) {\sim} -i A(\lambda ,z)
e^{-\alpha(\lambda ,z)}(1 + \dots ) \nonumber \\
& & \qquad + S\left[ \alpha(\lambda ,z) \right]A(\lambda ,z)(1 +
\dots ) e^{\alpha(\lambda ,z)}.
\end{eqnarray}
In the right-hand side of (\ref{AsympDebyeICORR}), the first term
is the usual Debye asymptotic expansion truncated near its least
term. The second one is obtained by decoding the divergent tail of
that asymptotic expansion. This can be done (see
Refs.~\onlinecite{Berry89,BerryHowls90}) by Borel summation after
exploiting a resurgence formula discovered by Dingle
\cite{Dingle73}. In the region of the $\lambda$-complex plane
where the Regge pole $\lambda_\mathrm{SP}$ lies (see
Fig.~\ref{fig:StokesH1}), we have $\mathrm{Re}~\alpha < 0$. As a
consequence, the first term of the right-hand side of
(\ref{AsympDebyeICORR}) is the dominant contribution while the
second one is a subdominant term which can be forgotten when
$|\lambda| \to \infty$. That is what we did previously by using
(\ref{AsympDebyeIa}). The Stokes multiplier function $S\left[
\alpha(\lambda ,z) \right]$ is a complicated function involving
the exponential integral function $E_1$. It goes continuously from
$0$ to $1$ at the crossing of the Stokes line $\mathrm{Im}~\alpha
= 0$ emerging from the turning point $z=\beta$ (see
Fig.~\ref{fig:StokesH1}). Below the Stokes line, it rapidly
vanishes. On the Stokes line, it is equal to $1/2$ and above the
Stokes line it rapidly becomes equal to $1$. It thus describes the
rapid but continuous birth of the subdominant contribution near
the Stokes line.

From (\ref{AsympDebyeICORR}) we can now write
\begin{eqnarray} \label{lhsRPSPcoor}
&  &\frac{1}{\sqrt{\epsilon_h}}
\frac{H_{\lambda_\mathrm{SP}}^{(1)'}(\sqrt{\epsilon_h} a\omega/c
)}{H_{\lambda_\mathrm{SP}}^{(1)}(\sqrt{\epsilon_h} a\omega/c )}
\sim -  \frac{\left[ \lambda_\mathrm{SP}^2 - \epsilon_h
(a\omega/c)^2 \right]^{1/2}}{\epsilon_h (a\omega/c)}  \nonumber \\
& &  \qquad  \times \left( 1- 2i S\left[
\alpha(\lambda_\mathrm{SP} , \sqrt{\epsilon_h} a\omega/c) \right]
e^{2 \alpha(\lambda_\mathrm{SP},\sqrt{\epsilon_h} a\omega/c)}
\right)  \nonumber \\
&  &
\end{eqnarray}
instead of (\ref{lhsRPSP1}). By using (\ref{rhsRPSP1}) and
(\ref{lhsRPSPcoor}), Eq.~(\ref{RPSP1}) can be solved approximately
and we find
\begin{subequations}
\begin{eqnarray} \label{RPSPasymBallO1}
&  & \mathrm{Re} \, \lambda_\mathrm{SP} (\omega ) \sim  \left(
\frac{\omega a}{c} \right) \sqrt{  \frac{\epsilon_h
\epsilon_c(\omega )}{\epsilon_h +
\epsilon_c(\omega )}},  \label{RPSPasymBallOa}  \\
&  & \mathrm{Im} \, \lambda_\mathrm{SP}(\omega) \sim 2
\left(\frac{\omega a}{c} \right)  P(\omega )S(\omega ) e^{2
\alpha(\omega)} \label{RPSPasymBallOb}
\end{eqnarray}
\end{subequations}
where
\begin{subequations}
\begin{eqnarray} \label{RPSPasymBallO2}
&  & P(\omega )\approx \frac{\epsilon_h^2 \epsilon_c^2(\omega
)}{\left( \epsilon_h^2 - \epsilon_c^2(\omega ) \right)
\sqrt{\epsilon_h \epsilon_c(\omega )(\epsilon_h +
\epsilon_c(\omega ))}} ,  \label{RPSPasymBallOc}  \\
&  & \alpha(\omega ) \approx \alpha \left((\omega a/c) \sqrt{
\frac{\epsilon_h \epsilon_c(\omega
)}{\epsilon_h + \epsilon_c(\omega )}},\sqrt{\epsilon_h} a\omega/c \right) ,  \label{RPSPasymBallOd}  \\
&  & S(\omega ) \approx S\left[ \alpha \left((\omega a/c) \sqrt{
\frac{\epsilon_h \epsilon_c(\omega )}{\epsilon_h +
\epsilon_c(\omega )}} , \sqrt{\epsilon_h} a\omega/c \right)
\right]. \nonumber \\
&  & \label{RPSPasymBallOe}
\end{eqnarray}
\end{subequations}
We can see easily that the imaginary part (\ref{RPSPasymBallOb})
of $\lambda_\mathrm{SP}$ vanishes for $\omega =\omega_s$ as well
as in the large radius limit $a\to \infty$, i.e. in the flat
interface limit. Furthermore, we have numerically studied
(\ref{RPSPasymBallOb}) for ``high" values of $\omega$, i.e. when
the Regge pole $\lambda_\mathrm{SP}$ is very close to the Stokes
line. In that case, by giving to the Stokes multiplier function
$S$ the value $1/2$, we have checked that (\ref{RPSPasymBallOb})
provides accurate results for the imaginary part of
$\lambda_\mathrm{SP}$. We therefore consider we have succeeded in
providing an analytic formula for the Regge pole of the SP.
However, the expression (\ref{RPSPasymBallOb}) is a rather
complicated function of $\omega $. As a consequence, its use in
the semiclassical formula (\ref{sc2}) is unfortunately not very
interesting. In short, we think that (\ref{RPSPasymBallOb}) is
especially interesting for the qualitative description of the SP
damping it provides.

\section{Conclusion and perspectives}

In this article we have introduced the CAM method in the context
of scattering of electromagnetic waves from metallic and
semiconducting cylinders. This allows us to provide a physical
explanation for the excitation mechanism of RSPM's as well as a
simple mathematical description of the surface wave (i.e. the SP)
that generates them. It should be noted that our results are not
limited to metals and semiconductors. Under simple assumptions,
they are also valid, {\it mutatis mutandis}, for more general
materials. Indeed, in a frequency range where the dielectric
function of a material presents a dominant simple pole $\omega_o$,
it is always possible to write \cite{MarkFox}
\begin{equation}
\epsilon_c (\omega) \thickapprox \epsilon_\infty + \frac{2
\omega_o R }{\omega_o^2 - \omega (\omega  +i \gamma)}.
\label{PermModLor}
\end{equation}
Here $\omega_o$ is the resonance frequency of the material in the
frequency region considered, $\gamma$ denotes the associated
damping term, $\epsilon_\infty$ stands for the high-frequency
limit of the dielectric function and we assume that the
coefficient $R$ is positive. In the absence of dissipation
($\gamma \thickapprox 0$) and if the zero of $\epsilon_c (\omega)$
which is given by $\sqrt{\omega_o^2 + 2 \omega_o
R/\epsilon_\infty}$ lies in the validity range of
Eq.~(\ref{PermModLor}), there exists a SP which can be described
by the theory developed in Sections 3 and 4.

In parallel with the semiclassical analysis of SP's on metallic or
semiconducting cylinders, we have developed a new picture of the
photon-cylinder system: it can be viewed as an artificial atom for
which the photon plays the role of an electron. RSPM's are
long-lived quasibound states for this atom, the associated complex
frequencies are Breit-Wigner-type resonances while the trajectory
of the SP which generates them is a Bohr-Sommerfeld-type orbit.
Furthermore, the imaginary part of a given RSPM complex frequency
corresponds to an exponentially small term which lies beyond all
orders in perturbation theory. As a consequence, as their
excitation frequency increases, RSPM's gain very long lifetimes,
i.e. they behave like bound states.

With in mind applications in photonic crystal physics, our work
could be naturally extended in various directions including i)
scattering by cylinders with metallic or semiconducting coating,
or more generally, multilayered structures, ii) scattering by
metallic or semiconducting spheres and last but not least iii)
scattering by objects fabricated from left-handed materials. It
would be also interesting to provide a complete (i.e., not limited
to SP's) semiclassical description of scattering of
electromagnetic waves from metallic and semiconducting objects in
the framework of CAM techniques by extending the ideas of
Nussenzveig and coworkers\cite{Nus92}. But at first sight that
seems to be a formidable task.

In recent papers dealing with photonic band structure of
two-dimensional photonic crystals fabricated from metallic or
semiconducting cylinders arrayed in a square lattice, a striking
feature has been noted \cite{McGurnMaradudin93,KusmiakMP1994,
KusmiakMG1997,KusmiakM1997,Moroz2000,Sakoda2001b,
MorenoEH2002,OchiaiSD2002}, namely the existence of flat bands
(i.e. dispersionless bands) in the frequency range in which
$\epsilon (\omega) <0$. This result, which only exists for $H$
polarization, is of course linked to the excitation of RSPM's.
More precisely, it is due to the localization of the photon which
is trapped on the Bohr-Sommerfeld orbit. Of course, this analysis
is rather over-simplified. In fact, it is necessary to understand
up to what point single-cylinder resonant aspects are related to
``resonant" aspects of the full photonic crystal. Recently, Ito
and Sakoda \cite{Sakoda2001b,Sakoda2001} have considered this
problem by developing a physically intuitive but appealing
analysis: they regard the RSPM's of an isolated cylinder as atomic
orbitals and they describe their effects into the photonic crystal
in the context of the ``linear combination of atomic orbitals
(LCAO) theory". Ito and Sakoda do not use the terminology
``artificial atom" to describe the photon-cylinder system but this
picture is implicitly present in their work and the point of view
we develop in the present article strengthens their analysis.

Of course, it should be interesting to provide a more rigorous
interpretation of the existence of the flat bands. With this aim
in view, it would be possible to benefit from the machinery
developed in semiclassical physics (see, for example,
Ref.~\onlinecite{BrackBhaduri} and references therein) to analyze
quantum chaos in connection with multiple scattering. As far as we
know, such a semiclassical approach has never been considered in
the context of photonic crystal physics but it seems to us very
promising. Indeed, it is well-known that, due to convergence
problems, band structure computations of metallic or
semiconducting photonic crystals are very heavy in the frequency
range in which $\epsilon (\omega) <0$. The semiclassical approach
could permit us to easily construct these band structures by
taking into account the shortest periodic orbits involving SP
trajectories and lying into the Wigner-Seitz cell. Here, the
properties we have found for the SP would be very useful.

\begin{acknowledgments}
We are grateful to Nick Halmagyi for help with English.
\end{acknowledgments}

\bibliography{CAMandSP-PRB}%
\end{document}